\documentstyle[11pt,epsf]{article}
\def\fnote#1#2{\begingroup\def\thefootnote{#1}\footnote{#2}\addtocounter
{footnote}{-1}\endgroup}

\def\beq{\begin{eqnarray}}
\def\eeq{\end{eqnarray}}
\def\bea{\begin{eqnarray*}}
\def\eea{\end{eqnarray*}}

\def\NPB#1#2#3{Nucl. Phys. {\bf B#1}, #3 (#2)}
\def\PLB#1#2#3{Phys. Lett. {\bf B#1}, #3 (#2)}
\def\PLBold#1#2#3{Phys. Lett. {\bf #1B}, #3 (#2)}
\def\PRP#1#2#3{Phys. Rep. {\bf #1}, #3 (#2)}
\def\PRD#1#2#3{Phys. Rev. {\bf D#1}, #3 (#2)}
\def\PRold#1#2#3{Phys. Rev. {\bf #1} (#2) #3}
\def\PRL#1#2#3{Phys. Rev. Lett. {\bf #1}, #3 (#2)}
\def\PREP#1#2#3{Phys. Rep. {\bf #1} #3, (#2)}
\def\ZPC#1#2#3{Z. Phys. C {\bf #1}, #3 (#2)}

\def\centeron#1#2{{\setbox0=\hbox{#1}\setbox1=\hbox{#2}\ifdim
\wd1>\wd0\kern.5\wd1\kern-.5\wd0\fi
\copy0\kern-.5\wd0\kern-.5\wd1\copy1\ifdim\wd0>\wd1
\kern.5\wd0\kern-.5\wd1\fi}}
\def\ltap{\;\centeron{\raise.35ex\hbox{$<$}}{\lower.65ex\hbox{$\sim$}}\;}
\def\gtap{\;\centeron{\raise.35ex\hbox{$>$}}{\lower.65ex\hbox{$\sim$}}\;}
\def\gsim{\mathrel{\gtap}}
\def\lsim{\mathrel{\ltap}}

\def\doublespaced{\baselineskip=\normalbaselineskip\multiply
    \baselineskip by 200\divide\baselineskip by 100}
\def\singleandhalfspaced{\baselineskip=\normalbaselineskip\multiply
    \baselineskip by 150\divide\baselineskip by 100}
\def\singleandabitspaced{\baselineskip=\normalbaselineskip\multiply
    \baselineskip by 120\divide\baselineskip by 100}
\def\singleandthirdspaced{\baselineskip=\normalbaselineskip\multiply
    \baselineskip by 130\divide\baselineskip by 100}
\def\singlespaced{\baselineskip=\normalbaselineskip}

\newcommand{\newc}{\newcommand}
\newc{\qbar}{{\overline q}}
\newc{\Kahler}{K\"ahler }
\newc{\deltaGS}{\delta_{\rm GS}}
\begin{document}
\begin{titlepage}
\begin{flushright}
{\large hep-th/0309170 \\ SCIPP-2000/04\\
}
\end{flushright}

\vskip 1.2cm

\begin{center}

{\LARGE\bf Is There A String Theory Landscape?}

\vskip 1.4cm

{\large  T. Banks, M. Dine and E. Gorbatov}
\\
\vskip 0.4cm
{\it Santa Cruz Institute for Particle Physics,
     Santa Cruz CA 95064  } \\

\vskip 4pt

\vskip 1.5cm

\begin{abstract}

We examine recent claims of a large set of flux compactification
solutions of string theory.  We conclude that the arguments for
AdS solutions are plausible.  The analysis of meta-stable dS
solutions inevitably leads to situations where long distance
effective field theory breaks down. We then examine whether these
solutions are likely to lead to a description of the real world.
We conclude that one must invoke a version of the anthropic 
principle which insists on carbon based life forms.
We explain why it is likely that this leads to a 
prediction of low energy supersymmetry breaking, but that many 
features of anthropically selected flux compactifications are 
likely to disagree with experiment.

\end{abstract}

\end{center}

\vskip 1.0 cm

\end{titlepage}
\setcounter{footnote}{0} \setcounter{page}{2}
\setcounter{section}{0} \setcounter{subsection}{0}
\setcounter{subsubsection}{0}

\singleandthirdspaced

\section{\bf Introduction}

String Theory has always been plagued by a plethora of solutions
which do not describe the real world.  Early hopes that some of
the highly supersymmetric (SUSic) moduli spaces of solutions would
prove inconsistent once nonperturbative physics was understood,
were dashed in a definitive manner by the second superstring
revolution.  More recently there have been claims
\cite{bp,sethietal,becker,sethib,gkp,acharya,kklt} that there is a large set of
additional solutions of string theory, both SUSic and SUSY
violating, with a {\it discretuum} of values of the cosmological
constant. Susskind\cite{landscape} has dubbed this discrete set of
solutions the {\it landscape} of string theory, and suggested that
anthropic arguments will have to be invoked to explain why the
world we see is chosen from among this vast class of solutions.
Douglas\cite{douglas}, has carried through what might be described
as a prototype counting of such
vacua.  Ultimately, one might
hope determine the distribution of couplings among the states of the
landscape (what Douglas
refers to as the ``ensemble'').

The purpose of this paper is to assess the arguments for the
existence of the flux discretuum, as well as the prospects that
any of these solutions could describe experimental physics. Our
conclusions may be summarized as follows: the arguments for the
existence of AdS solutions, both SUSic and SUSY violating, appear
plausible.  The key to verifying their existence would be the
discovery of dual CFT's for each of these solutions.
Silverstein\cite{silverstein} has suggested a method for constructing
these duals using brane/flux transitions. The meta-stable dS
solutions are on much less firm ground.  Many, perhaps all, of
them are likely to tunnel to Big Crunch cosmologies, where the low
energy effective field theory technology, which is the primary
tool for constructing these solutions, breaks down. We argue that
these instabilities must be faced even if the amplitudes for them
are much smaller than those for other decay modes.  Furthermore,
even in the absence of these instabilities, the putative stable
state into which these solutions decay, is a Big Bang universe
where effective field theory breaks down. Unless we develop
techniques for understanding these singularities, there is no
reliable evidence that these solutions are approximations to
meaningful models of quantum gravity.

The existence of a discretuum of stable or metastable states would
force us to adopt a new paradigm for scientific
explanation\footnote{We are assuming, here, that there is not some
sort of (possibly cosmological) selection principle which picks
out one or a few states; in this case, there is not really a
discretuum of states.  An example of such a principle is the
hypothesis of cosmological SUSY breaking\cite{Banks:2000fe}.}. We will discuss the
possible paradigms, but, as we will explain, at least within the
classes of flux vacua which have been considered up to now, the
possibility that the real world is described by one of these
states appears somewhat dim.

One possible mode of explanation is to simply take a set of
observed experimental facts, the gauge group, particle content and
couplings of the Standard Model, the spectrum of fluctuations of
the CMBR and the like, as given, and ask whether further
predictions are possible.  The second is to adopt an anthropic
explanation.  In the first case, as we will explain, it is
essentially impossible to determine which is the ``true vacuum"
within the discretuum. It is necessary
to ask what (if any) subset of states is consistent with the
observed experimental facts and whether the distribution of some
as yet unmeasured quantities is highly peaked about some values (this
viewpoint has been stated most forcefully by Douglas).
The flux discretuum would appear to be a natural
framework in which to implement the anthropic
principle\cite{anthropic}.
Indeed, it is clear that in the context of the landscape, this
is the only way in which to {\it explain} the value of the
cosmological constant, and probably several other quantities of
low energy physics.  But we will see that in the flux discretuum,
such a viewpoint will almost inevitably lead to predictions which
are false.

We have previously presented \cite{bdm} a variety of arguments
against the utility of the anthropic principle in models of this
type. In asking about anthropic vacuum selection, we distinguished
two possibilities (what Dimopoulos has dubbed ``habitability
criteria"), which we will refer to as 
``Selection Principle A" and ``Selection Principle B".  Principle
B
selection principle asserts that any states which meet some very
minimal criteria for the existence of life -- the formation of
structure in the universe, perhaps of stars, and the like, are
acceptable. Selection principle A postulates the
necessity for our own type of carbon based life.  Principle
B is a very reasonable constraint on
mathematical theories of the universe, but is not very powerful.
Principle A is potentially much more powerful, but requires one
make assumptions which are very difficult to justify.
Given a collection of low energy gauge groups and representations,
we simply do not understand enough about complex phenomena in (the
various possible analogs of )low energy nuclear physics, much less
(exo)biology, to assert that life is or is not possible.
One of our points will be that even if we implement some form of principle
A, the landscape is likely to make many incorrect predictions.

In \cite{bdm}, we presented a variety of arguments against the
utility of even principle A in models of the
class exemplified by the flux discretuum.
In this paper we modify and extend those arguments, explaining why
the predictions of principle A within the
framework of these models , are likely to differ widely from
experimental values of many parameters.

Apart from these negative comments, we note that there is one
interesting prediction that the flux vacuum might make:  low
energy supersymmetry.  In particular, we explain why the vast
majority of vacua with cosmological constant and
gauge hierarchy required by principle A, are likely to exhibit some form of low energy
supersymmetry.  This is not necessarily a success; again, within
the framework of principle A, our
arguments suggest that the flux discretuum is not likely to agree
with experiment.  Some, though not all, of these observational
problems are linked to the prediction of low energy SUSY.

We begin our discussion in the next section by critically
examining the notion of the low energy effective Lagrangian, and the
assumption that different solutions of the same long wavelength
effective field equations are related to the same model of quantum
gravity.  Section 3 is devoted to a discussion of AdS and
(primarily) meta-stable dS solutions of long distance effective
SUGRA, presenting our detailed argument for the conclusions
adumbrated above.  In Section 4 we turn to phenomenological
questions and the anthropic principle.  Section 5 reiterates our
conclusions.  The system behind our exposition is to present a
series of potential problems with the notion of a string theory
landscape.   These problems are ordered from fundamental to
phenomenological.  At each stage of the exposition, we ignore the
objections of the previous stage, but the reader is certainly
expected to keep them in mind.

\section{\bf The proper use of effective lagrangians}

To explain our point of view about effective lagrangians in string
theory, we will have to introduce some non-standard terminology.
We will use the term {\it long wavelength effective lagrangian} to
describe the object that is conventionally derived from quantum
field theory or string S-matrices.  In \cite{hetero}, one of the
authors (TB) argued that, when supplemented with asymptotic
boundary conditions in space-time, the long wavelength effective
lagrangian in a model of quantum gravity also carries information
about the high energy spectrum of the theory, via the properties
of its black hole solutions.  It was also argued that there were
disjoint moduli spaces of models of quantum gravity, which could
not be thought of as different states of the same theory.   The
latter notion (that of superselection sector in quantum field
theory) depends on a separation between long wavelength and high
energy physics, which is simply not valid in models of quantum
gravity.  We will use the term {\it model of quantum gravity} to
refer to a single connected moduli space of S-matrices or boundary
correlation functions\footnote{If there is a model describing
stable asymptotically dS space, then it has a finite number of
states and no precisely defined observables.  Nonetheless, we
would like to include this possibility in our list of possible
models.}. The term {\it theory of quantum gravity} will be used to
refer to an as yet only partially understood set of rules for
constructing all sensible moduli spaces of models. The terms {\it
String Theory} or {\it M-theory} should be understood in this
sense. We believe that there are many mathematically consistent
models of quantum gravity, at most one of which describes the real
world.

The main point of this section is that the same long wavelength
lagrangian can describe the dynamics of different models of
quantum gravity.  This is best explained by examples.  Consider
the model called ``M-theory Compactified on $T^d$", with $0 \leq d
\leq 7$.  It is defined by an S-matrix in $11-d$ asymptotically
flat dimensions. The low energy dynamics of this model can be
described by a long wavelength effective lagrangian.  The
lagrangian can be calculated by matching scattering amplitudes
computed with it to amplitudes computed from the fundamental
formulation of the theory.   In terms of the lagrangian this
corresponds to solving the classical equations of motion with
scattering data of infinitesimal strength.  The S-matrix is
computed in terms of the asymptotic expansion of the solution in
the strength of the scattering data.  The effective action point
of view is useful for computing the S-matrix as an asymptotic
series in various ratios of length and energy scales.

The same lagrangian has classical solutions corresponding to $11$
dimensional FRW cosmologies.  In these solutions, the moduli of
M-theory on a torus can vary with cosmological time.  Spatial
sections of the cosmology can have positive, negative or zero
scalar curvature. It is clear that the observables (the S-matrix)
of the well understood model do not generalize to these
cosmological solutions.  For any cosmological solution, the past
or future asymptotic region \footnote{or both} is replaced by a
singular space-time where all known approximate methods for
solving M-theory break down.  It may be that some of these
cosmological solutions represent aspects of the physics of a well
defined model of quantum gravity {\it different from M-theory
compactified on a torus}.  At present we do not have the tools to
study this question, or to decide whether various cosmological
solutions are all related to the same model.  It seems likely that
at least some of these solutions do not correspond to any sensible
model. Thus, the existence of M-theory compactified on a torus as
an S-matrix theory in asymptotically flat space, does not
guarantee the existence of these cosmological models.

A similar example can be generated by studying M2-branes near
conical singularities, whose near horizon geometry has the form
$AdS_{4} \times K$, with $K$ a compact Einstein manifold.  The
geometrical moduli of $K$, become coupling constants in the family
of CFT's which can be obtained in these limits.  Each is a
different model of quantum gravity. By studying nearly singular
limits of $K$ we can generate examples of this type with a single
light supermultiplet, much lighter than any of the other massive
states of the theory. It is believed that the exact quantum
dynamics of this system is described by a conformal field theory
in $2+1$ dimensions.  The long wavelength effective lagrangian is
SUGRA coupled to the light scalar multiplet, with a negative
cosmological constant. We can find other solutions of the same
lagrangian, among them matter dominated FRW universes with Big
Bang and/or Big Crunch singularities, as well as the Godel
universe.   Again it is clear that these are not all part of the
same model of quantum gravity. The existence of the exact CFT
description of the quantum dynamics of the AdS solution says
nothing about the existence of quantum models which might be
approximated by these other solutions.

The lesson of these examples is plain.  Quantum gravity is not
like quantum field theory, where a change in boundary conditions
changes only a limited set of degrees of freedom of the system.
In gravitational physics the asymptotic boundary conditions
define the whole system.  Our idea that different solutions of a
long wavelength effective lagrangian are really part of the same
theory is a relic of QFT intuition, and must be discarded. It
follows that demonstrating the existence of an extra solution to a
lagrangian derived from one well defined model of quantum gravity
does not by itself imply that the solution corresponds to another well
defined model.

These examples are consistent with the message of
\cite{hetero,isovac}, which is that the off shell effective
potential is not a meaningful quantity in models of quantum
gravity.  It can at best be viewed as a heuristic tool, valid in
some extreme regions of moduli space.

\subsection{The Effective Potential in String Perturbation Theory}

The arguments of \cite{hetero,isovac} and those based on
AdS/CFT, depend on properties of black holes, which are
non-perturbative in the string coupling.  This might lead one to
suppose that considerations based on the effective potential
might be well defined to all orders in string perturbation theory
when vacuum energies are of order the string scale or less. There
is no evidence that this is the case.   The effective potential,
when it is non-zero, is an off shell quantity and we generally
have no way to compute it beyond the first order in which it
appears.

For many years, it was believed that the work of Fischler and
Susskind\cite{fs} provided a method for computing systematic
corrections to the effective equations of motion in string
perturbation theory.   The idea was that divergences in integrals
over the moduli space of Riemann surfaces would, for a general
tree level string background, contribute anomalous terms to the
BRST Ward identities for higher genus string computations:
\beq
{\sum_i <V_1 \ldots [Q, V_i] \ldots V_n >_g \neq 0}
\label{brstanom}
\eeq
A non BRST invariant perturbation of the world sheet lagrangian,
of order $g_S^{2g-2}$ would be inserted to restore BRST
invariance of the full sum of amplitudes, up to that genus.

This prescription appears to provide a way of correcting the
equations of motion of fields which are massless at tree level, to
all orders in string perturbation theory.  Of course, these
manipulations do not define an off-shell potential.  They merely
enable one to obtain quantum corrections to the on-shell
condition. For massive fields, there is no analogous
prescription, and we might not expect one. Most of these fields
correspond to unstable excitations, which do not appear in the
S-matrix.  Even for stable massive modes, the possibility of
doing nonlocal field redefinitions, with non-locality of scale
$\sqrt{\alpha^{\prime}}$, makes the notion of effective potential
ambiguous.  As a consequence it is not at all clear how to use
these perturbative methods to compute the properties of minima at
finite values of the string coupling, as would be needed to discuss
flux compactifications.

Even for massless fields,
there is a technical
problem\footnote{Pointed out to us by Eva Silverstein.} with the
Fischler-Susskind program.   The one loop BRST anomaly produces a
tadpole for the dilaton field.   The linearized solution of the
corrected dilaton equation of motion grows like two powers of the
time.   However, verification of the BRST Ward identity involves
a partial integration in space-time.   Thus, there is a
space-time surface term in the BRST anomaly, which does not
vanish even when we solve the Fischler-Susskind equations.

This might be attributed to a failure to solve the full nonlinear
equations of motion.  However in \cite{bdcosmo} two of the present
authors argued that nonlinear solutions to the FS equations
always replaced at least one of the two asymptotic boundary
conditions with a finite time space-like singularity.   Thus, the
FS program does not give a controlled expansion of the string
equations of motion.  The small perturbative corrections become
arbitrarily large at late or early times.  We emphasize that this
all occurs in a region where the string coupling is weak.

To summarize, even in perturbative string theory, we have no
systematic definition of an effective potential.    At this stage
an analogy might be useful.   When computing the amplitude for a
quantum system to move between two points in configuration space
over some time interval, one can often apply the WKB
approximation, and do the computation in terms of a classical
solution of the equations of motion with boundary conditions that
it hits the two points at the ends of the given time interval. It
is certainly possible to calculate quantum corrections to the
amplitude, but there is no known meaning to the quantum
correction to this classical path\footnote{There is a quantum
effective action, which is a generating functional for Green's
functions.   One can compute quantum corrections to the action
and find modified ``classical" solutions of it.   But their
purpose is to be expanded in powers of a source and generate
Green's functions.  They do not modify the path between a pair of
points.}.  The next correction to the amplitude sums over {\it
all} deviations from the classical path.

The point of this analogy is that the effective potential in
theories of quantum gravity may play a role analogous to the
classical path between two points in ordinary quantum mechanics.
In leading approximation it may be a device for determining the
stability of a hypothetical model of gravity, but there may be no
exact object in the theory to which it is an approximation. Much
of the thinking implicit in discussions of flux compactifications
depends on the notion that the effective potential is an exact
object, for which we are presently able to find only approximate
expressions.   This line of thought might be completely wrong. We
have no evidence from string theory or gravity that such an
object exists.

In the remaining sections of this paper, we will ignore the
caveats we have just enunciated.  That is, we will explore
hypothetical solutions to string theory which are motivated by
finding minima of a low energy effective potential.   The
thoughtful reader will however, keep these remarks in mind as he
follows our path through the hypothetical landscape of string
theory.

\section{\bf Exploring the Landscape of Flux Compactifications}

For definiteness, we will restrict our attention to
compactifications of six dimensions in Type IIB string theory,
which have been explored in great detail.  One begins by searching
for SUSic flux compactifications, using the GVW\cite{gvw}
superpotential in the presence of fluxes, branes and
orientifolds\footnote{A naive person might have imagined that it
was inconsistent to include D-branes, which are singular sources
with energy of order ${1\over g_S}$ in string units. However, for
SUSY preserving solutions, all terms of this order cancel exactly.
For SUSY violating compactifications this issue must be revisited.
We are not sure of the status of this issue at present. } . The
fluxes are quantized, and the stationarity condition for the
superpotential fixes the six manifold to be conformally
Calabi-Yau, as well as fixing the dilaton and the complex
structure moduli.    In the classical approximation the Kahler
moduli are left free because the superpotential does not depend on
them.   We obtain a no-scale SUSY breaking solution with vanishing
cosmological constant.   It is clear that corrections to the
Kahler and super potentials will drastically change this scenario.
The authors of \cite{kklt} study a modification of the
superpotential due to gaugino condensation, in models with only
one Kahler modulus. This induces dependence of the superpotential
on the Kahler modulus, and allows for a SUSY
preserving minimum where the Kahler modulus is fixed and the
cosmological constant is negative. This is the first indication of
a landscape.  By using the wide variety of fluxes available on a
$CY_3$ with large $b_3$ we can construct a plethora of SUSic AdS
solutions of the effective field equations.

We consider this construction to be highly plausible, although
there is no sense in which these solutions are part of the same
model of quantum gravity as the solutions with vanishing flux and
vanishing cosmological constant.  There is nothing in
the physics of these solutions which indicates a breakdown of
long wavelength effective field theory (as long as the fluxes are
chosen to be sufficiently large that all dimensions of the
manifold are large compared to the string scale).  Even effective
field theory corrections to the cosmological constant are under
control because of exact SUSY. There are only two disturbing
aspects of this construction. Although much effort goes into guaranteeing that
the string coupling is small, there is no systematic string
perturbation theory around these solutions.  Indeed, although we
have invoked non-perturbative low energy physics to stabilize the
Kahler modulus, there is no rigorous way of estimating the
corrections, let alone a systematic machinery for calculating
them.   More generally, since for fixed topological numbers, the
number and magnitude of fluxes that are allowed for
compactifications to 4 dimensions are bounded from above, there
is actually no expansion parameter which could be made
arbitrarily small\footnote{The question of bounds on the
flux number is somewhat subtle.  The usual constraint arises
from tadpole cancellations.  In this relation, it would appear that
by suitable choices of signs, some fluxes could be made arbitrarily
large.  As we will see, however, in these limits, the approxmimations
of \cite{kklt} break down.}

For the case of solutions with negative cosmological constant, a
truly rigorous test of these proposals would be to discover the
three dimensional CFT's which are dual to them. Progress along
these lines was reported by Silverstein\cite{silverstein} at the
Strings 2003 Conference in Kyoto.  The basic idea is to use
``flux-brane'' transitions\cite{fluxbrane} to convert a flux
compactification into a configuration of D-branes, and to study
the world volume field theory of the resulting branes.  It will be
a $2+1$ dimensional quiver gauge theory.  If the corresponding
$3+1$ dimensional theory is not asymptotically free, one can
imagine obtaining a perturbative zero of the $\beta$ function by
cancelling the running of the coupling from its engineering
dimension, with a large one loop quantum correction.   Such
perturbative fixed points cannot represent large radius AdS
compactifications.  However, in a similar situation in $3+1$
dimensions, the perturbative fixed points actually persist (at
least for SUSic theories) into the strong coupling regime. One
might hope that by varying the discrete parameters specifying the
putative quiver duals to large radius flux compactifications, one
could find perturbatively accessible fixed points whose existence
would lend support to the claim that the quivers indeed flow to fixed
point theories.

Thus, although the notion of off shell effective potential does
not have a meaning in models of quantum gravity, there are
indications that the prediction of a discretuum of AdS vacua,
which follows from this formalism, is plausible. Before
discussing dS vacua, we revisit the analysis of the effective
lagrangians which yield the supersymmetric AdS flux vacua.

\subsection{Detailed Analysis of the Discretuum}

 In this section, we will
review the analysis which suggests the existence of the flux
discretuum.  We want to focus on the question of the degree to
which the existence of the flux discretuum, even within the
framework of conventional effective field theory analysis, has
been established. The points we will make here have all been made
in \cite{kklt}, but they are worth emphasizing.  In particular, we
want to address the question of whether there is a small parameter
which could justify the analysis.

The authors of \cite{gkp} were interested in exhibiting warped
vacua of string theory in a more or less controlled approximation.
They found that, for fluxes located on a collapsing three cycle
near the conifold point, one does obtain such warping. They also
found that some of the moduli (the modulus associated with the
deformation of the conifold, in particular) were fixed.  In the
approximation in which they worked, the superpotential for this
modulus, $z$, along with the dilaton multiplet, $\tau$, was given
by: \beq W= (2 \pi)^3 \alpha^\prime (M {\cal G}(z) -N \tau z) \eeq
where \beq {\cal G}(z)= {z \over 2 \pi i}\ln(z) +
{\rm~holomorphic}. \eeq

This has a supersymmetric minimum where \beq D_z W = {\partial W
\over \partial z}+{\partial K \over \partial z}W =0 \eeq which is
solved by \beq z \sim exp(-{2\pi N \over M g_s}) \eeq If the ratio
$N/M$ is large, then $z$ is very small.  The corresponding space
can be shown to be highly warped.   The superpotential is
exponentially small at this stage.

They then noted that by including additional fluxes, it is
possible to fix other complex structure moduli, including $\tau$.
The resulting superpotential
has the structure:
\beq
W= (2 \pi)^3 \alpha^\prime[M {\cal G}(z)-\tau (Kz + K^\prime f(z))]
\eeq
This has a supersymmetric stationary point, $$D_\tau W = {\partial W \over \partial \tau}
+ {\partial K \over \partial \tau} W =0$$
when
$$\bar \tau = {M {\cal G}(0) \over K^\prime f(0)}~~~~~W= 2(2 \pi)^3 \alpha^\prime
M {\cal G}(0) $$

The resulting stationary point still yields a warped geometry, but
now at the stationary point, $<W>=W_0$ is no longer exponentially
small; it is large in Planck units; in fact, including fluxes to
fix the other complex structure moduli, its value in Planck units
is typically of order some large flux number.  We will see however
that special configurations can have relatively small $W_0$.  We
believe that at best, for reasons which we will now explain,
the analysis can be justified only for these special
states.

It should be noted here that in the best of cases there is no
systematic computation which justifies this calculation.  The flux
numbers cannot be taken arbitrarily large, for fixed $CY_3$.
Still, the fact that these numbers {\it can} be quite large in
particular examples creates the hope that the results are
meaningful.

In this approximation, the radial modulus is not fixed.   The
authors of \cite{kklt} argued that various processes, such as
gaugino condensation in the low energy theory, would modify the
superpotential, adding terms of the form \beq W = W_0 + e^{i\rho/b}
\eeq where $\rho$ is the radial multiplet, and $b$ is a constant.
At the stationary point, $e^{i\rho/b} \sim W_0$.  But we have noted
that $W_0$ is typically large, so $\rho$ is small, and the
analysis is not self consistent.  The spirit of \cite{kklt} is
that, given the huge number of possible choices of
flux, $W_0$ will be small for many choices.  For these, $\rho$
will be large.  In addition, while the masses of the Kaluza-Klein
modes are of order $1/R^2$, the masses of the dilaton and the
complex structure moduli are of order $1/R^3$ (though possibly
times large flux factors), and the mass of $\rho$ will be of order
$\vert W_0 \vert^2 \over R^2$.  So the picture of first
integrating out higher dimensional physics, followed by the
complex structure moduli, followed by $\rho$, is consistent.

We should note, however, a potential difficulty.  The masses of
the complex structure moduli are typically enhanced by a factor of
flux-squared, i.e. they are of order $N^2 \over R$ relative to the
Kaluza-Klein masses.  So one requires {\it very large $R$} if
these are to be small relative to the Kaluza-Klein scale.  More
precisely, this ratio is of order \beq {m_{cs}^2 \over m_{kk}^2}
\sim {K^2 \over \rho \tau} \eeq where $K$ is a typical flux
number.  So if $\tau \sim K$, then we require $\rho \gg K$. Recall
that $e^{i b \rho} \sim W_0$. So $W_0$ must be extremely small.
For small $W_0$, the distribution of $W_0$'s might be expected to
behave as \footnote{Kachru (private communication)
has explained this heuristically by noting that, if
one thinks of the fluxes as forming a vector, $W_0 = \vec N \cdot
\vec a$, where $\vec a$ is a set of constants, which for large
$\rho$ are roughly independent of $\rho$.  So $W_0$ can be thought
of as the dot product of two vectors in a high dimensional space;
calling the azimuthal angle $\theta$, one is interested in the
distribution of $\cos(\theta)$ around $\theta= {\pi \over 2}.$
This picture cannot be taken too literally, since the the $a$'s, for example,
depend on the complex structure. }
\beq P(W_0) \propto {W_0 \over M_p^3} \eeq So only in an
exponentially small fraction of states is the analysis which leads
to the flux vacua completely under control.

It should be noted at this point that if one tries to make some fluxes
extremely large, consistent with the tadpole cancellation condition,
the low energy field theory approximation certainly breaks down.

The authors of \cite{kklt} went on to suggest one possible
mechanism for supersymmetry breaking.  They argued that if one
includes $\overline{D3}$ branes in these compactifications, they
would naturally sit near the (resolved) singularity, and would
have exponentially small tension due to warping.  Their presence
breaks the supersymmetry, so one would have a small breaking of
supersymmetry. The actual cosmological constant depends on an
interplay between $W_0$ (determined randomly) and the exponential
warp factor.  These are not correlated in any particular way, but
in some subset of states, one might expect (at least at the level
of the effective action) to find supersymmetry breaking with a
small, {\it positive} cosmological constant, as a result of cancellations.

There are other ways in which one might imagine supersymmetry
breaking to arise.  For example, the low energy theory might be a
theory in which supersymmetry breaks dynamically\footnote{This
raises puzzling issues, which will be dealt with
elsewhere\cite{bdginprog}}. Again, using conventional effective
field theory reasoning, one might expect to find,in some fraction
of vacua, a small, positive cosmological constant, as a result of
cancellation between physics at two, a priori unrelated, scales.

One question here is what the role of supersymmetry is in this
analysis.  After all, in some fraction of states, the radii might
be expected to be large and the cosmological constant small,
simply by accident. We will return to this question shortly and
discuss whether the flux discretuum might lead to a {\it
prediction} of low energy supersymmetry.

\subsection{Vacuum Tunneling and the Existence of Meta-stable dS States}

Conventional discussions\cite{bp,sethietal} of flux
compactifications envisage vacuum tunnelling between solutions
with different values of the cosmological constant.  This is often
assumed to be the mechanism which dynamically implements the
anthropic principle.  The universe jumps around between vacua
until it finds itself in an anthropically allowed one, at which
time we observe it.

In fact, as was recently emphasized in \cite{heretic}, the
original Coleman-De Luccia paper on vacuum tunnelling in the
presence of gravity implies important modifications to this idea.
These modifications are amply confirmed\cite{heretic}\cite{hetero}
by the modern interpretation of the quantum mechanics of AdS
universes in terms of conformal field theory.  We will first list
them and then explore their implications for flux compactification
models.

\begin{itemize}

\item There are no semiclassical tunnelling amplitudes into
AdS space times.  When there is a non-vanishing amplitude to
tunnel into the basin of attraction of a negative cosmological
constant minimum, the post-tunnelling classical evolution is a
negatively curved FRW cosmology, which undergoes a Big Crunch.

\item Amplitudes for tunnelling between any pair of
asymptotically dS universes, are finite.  This is true in the
sense that the post-tunnelling solution inside a cosmological
horizon volume, asymptotes to the dS space with appropriate
cosmological constant.  The global structure of the
post-tunnelling solution is not that of dS, but this may be
unimportant, since no observer can see more than a horizon volume.
The ratio of probabilities for jumping in the two directions is
governed by a law of detailed balance with the entropy rather than
the free energy in the exponent. This is consistent with
thermodynamics if most of the states of dS space have energies
below the dS temperature.

\item  There are no semiclassical tunnelling amplitudes between
dS space and asymptotically flat space.  When there is a
non-vanishing amplitude to tunnel into the basin of attraction of
a region of field space with vanishing cosmological constant, the
post tunnelling evolution is a negatively curved FRW solution,
which is not asymptotically flat.   It is often nonsingular in the
future, but has a Big Bang singularity in its past.

\end{itemize}

Thus, while the idea of jumping between dS states might make sense
in the semiclassical approximation, jumping to regions with
non-positive cosmological constant always involves a breakdown of
effective field theory.

We suspect that the Big Crunch singularities are a more serious
problem, and do not occur in well-defined models of quantum
gravity.  This suspicion is based on the AdS/CFT correspondence.
An asymptotically AdS universe is described quantum mechanically
by a conformal field theory, and the cosmological constant
controls the asymptotic density of states in the CFT.   Such a
theory is by definition stable, and has a conformally invariant
vacuum state.   Thus, although long distance effective Lagrangians
can exhibit tunnelling between AdS vacua and Big Crunch FRW
universes, this does not appear to be a feature of consistent
theories of quantum gravity.   Of course, this does not imply that
the analogous tunnelling solutions for non-negative cosmological
constant cannot occur, but it is suggestive.

If we suppose that tunnelling between dS minima can occur as
suggested by the semi-classical approximation, the above
discussion suggests a potential instability for the whole complex
of metas-stable dS flux vacua.   Suppose that in a consistent
model, one cannot tunnel into a Big Crunch.  If any meta-stable dS
state can tunnel to any other, and if one of them can tunnel to a
Crunch, then the whole system might be inconsistent.

The reader may object that if the amplitudes to tunnel to disaster
are much smaller than those to tunnel to a zero cosmological
constant region (which always exists in flux compactifications)
then they should be ignored in the semi-classical approximation.
We present the following quantum mechanical model as a
counter-example:  Consider the motion of a particle in a plane
under the influence of a potential

\beq
V(r,\theta ) = {(r^2 + a^2) \over (b^2 + r^4 )}  \cos^2 \theta + \epsilon r^4
\cos^5 \theta
\eeq
There is a meta-stable minimum at the origin, with two
semiclassical modes of decay.   Let us take $\epsilon \ll 1$. Then
there is a stable minimum at $\theta =0$ and $r \sim \epsilon^{-1/6}$.
The potential is unbounded from below when $\pi \leq \theta <
2\pi$ and the Hamiltonian does not exist as a Hermitian operator.
However, for small $\epsilon$ the tunnelling amplitude to the
unbounded region goes to zero exponentially, while that to the
stable minimum in the forward direction goes to a constant.  If
all we knew how to do was the semi-classical physics of this
model, we might have mistakenly ignored this sign of catastrophic
instability.  In ordinary quantum mechanics,  semiclassical
tunnelling calculations explore the potential surface.  Thus, even
when a given tunnelling amplitude is too small to be included in a
systematic expansion scheme, it may reveal true instabilities of
the system.\footnote{Susskind argues that the Big Crunch singularities are innocuous
because there are causal patches which are causally disconnected from
them.  Thus the singularities should be viewed as black holes from
the point of view of observers in these patches\cite{sussbang}}.

Even if this potential instability turns out to be a chimera, the
semiclassical discussion of meta-stable dS states still contains a
potential problem.   One cannot understand a meta-stable state
without understanding the stable system of which it is a temporary
excitation.   The only plausible candidate for a semiclassical
description of that stable state is, in the context of flux
compactifications, a negatively curved FRW solution with moduli
moving on the Dine-Seiberg region of the potential.   This is the
space-time that appears in the analytic continuation of the
instanton for meta-stable dS decay. There are good and bad aspects
of this suggestion.

It is often assumed, by analogy with quantum field theory, that a
meta-stable dS state could be an excitation of an asymptotically
flat space-time.   In field theory this is consistent.  The
post-tunnelling solution indeed asymptotes to flat space with
constant field at the stable minimum.   Furthermore we can excite
arbitrarily large regions of the stable flat space into the
positive energy density minimum.  Neither of these statements is
true when gravity is dynamical.  Guth and Farhi \cite{gf}
demonstrated that if one tries to excite a field coupled to
gravity into a meta-stable minimum of its effective potential, one
instead creates a black hole.   Consistent with this observation,
the post-tunnelling solution does asymptote to an FRW cosmology\footnote{If
the zero c.c. minimum is a stable minimum at a finite point, this
will be a matter dominated FRW model.  If, as is the case in
string theory, the minimum is at infinity, then we get a kinetic
energy dominated FRW with fields in the Dine-Seiberg region of
moduli space.} . Furthermore, it is at least plausible (given the
quantity of literature on old inflation) that a negatively curved
FRW space-time can, at sufficiently early times, be excited into a
meta-stable dS state. This question deserves much more careful
study, and is the heart of the issue.  To study it, one would have to
develop a quantum theory of negatively curved Big Bang FRW universes and
determine whether it really exhibited meta-stable dS states.

The real fly in the ointment is that the possibility of exciting
the meta-stable dS state depends precisely on the fact that the
FRW solution evolves from a Big Bang singularity.   That is, it is
because of the increase of energy density at early times that we
believe that we can easily go over the barrier to the meta-stable
state.  It would be disingenuous to ignore the fact that this is
happening because we are approaching a region where effective
field theory is breaking down.

We are quite sympathetic to the idea that Big Bang singularities
(or their proper resolution) may be part of {\it some} consistent models of
quantum gravity.  After all, our own universe appears to
have such a singularity in its past.   However, without a thorough development of the
full quantum theory of such systems, it seems premature to
conclude that the landscape of meta-stable dS minima is really a
feature of string theory.  It is only by finding a full quantum
description of the FRW solution that we can hope to calculate
corrections to the effective field theory description of the
meta-stable minima.

This brings up the question of how the meta-stable dS minima are
related to mathematically well defined observables  of the
putative stable FRW solution.   The asymptotic future of the
latter is a smooth space-time where asymptotic data can be
formulated.   Thus we can imagine that the quantum theory of such
a space-time is defined by an S-matrix or S-vector\cite{witsusbf}
relating such future asymptotic states to a basis of states
defined at the Big Bang.   The term S-vector refers to the
possibility that only one state is allowed at the Big Bang (or in
\cite{bfcosmo} that all states are gauge equivalent there). If
there is indeed an S-matrix, one can imagine picking out a
particular meta-stable dS state and tuning the scattering data to
find an S-matrix element or elements in which it appears as the
dominant resonance.   The relation of these S-matrix elements to
local measurements performed by meta-stable observers in the
meta-stable dS state is obscure, but presumably if we discover the
appropriate quantum theory we will be able to work it out.
\footnote{Susskind has suggested that we interpret the Big Bang of this solution
as the expanding phase of a Big Crunch-Big Bang solution.  The metastable
dS vacua appear as resonances in an S-matrix for this time symmetric
system.  In our opinion, one must resolve the singularity to understand
this system \cite{sussbang}.}

It is much less clear what to do if the system only has an
S-vector.  Then one cannot tune initial conditions and any
amplitude would be some complicated superposition of contributions
from the different dS states.

In summary, the landscape of meta-stable dS minima suffers from a
potentially catastrophic instability.  In addition, the candidate
for the stable state into which it decays suffers from a Big Bang
singularity.   One cannot conclude that the landscape exists
without understanding these singularities, not something
we can do with the effective field theory techniques we currently have at our
disposal.   One will have to find a more fundamental definition of
these states in order to calculate any of their properties beyond
the effective field theory approximation.  We emphasize that this
issue is connected with the question of parameters,
which we asked above.   In order to assess whether an expansion
parameter is small enough for us to believe the predictions of a
leading order calculation, we must have some understanding of the
nature of the exact model to which the leading order is an
approximation.  For meta-stable dS minima, this exact model would
be the theory of the S-matrix or S-vector in an FRW space-time
with negative spatial curvature and a Big Bang singularity.   It
is only when we have some clue to the correct formulation of the
quantum theory of such systems that we can really begin to assess
whether meta-stable dS minima exist.  Without such an exact
definition, we cannot compute even one step beyond the leading order
in the putative expansion.

In the remainder of this paper, we will assume that the landscape
exists and examine the question of whether it is likely to lead to
a description of the real world.  An immediate question to ask is
how the cosmological constant would be explained in such a
scenario.   Controllable examples are expected to have a cosmological
constant which is smaller than string scale, but not much smaller
than the scale of SUSY breaking.   However, the huge size of the
discretuum suggests the possibility that the mechanism of
\cite{bp}\footnote{An older variant of this idea appeared in
\cite{bds} .} for obtaining a tiny cosmological constant and a
large SUSY breaking scale might be operative.   This explanation
relies on the anthropic principle\cite{anthropic}.   In a huge collection of
meta-stable dS vacua, a minimum with a small value of the
cosmological constant might be picked out by the anthropic
principle.   In the context of flux compactifications, it is
implausible that there will be only a few minima satisfying the
anthropic constraints on the cosmological constant - many other
parameters of the long wavelength effective field theory are
likely to vary discontinuously as one jumps from one minimum with
an anthropically allowed c.c., to another.   Much of the
remainder of this paper will be concerned with the likelihood
that anthropic predictions for low energy physics actually fit
the data.

\section{Scientific Explanation in the Discretuum}

There are two possible modes of scientific explanation in the
discretuum. The first is to simply to ask whether there are states
in the discretuum with properties identical to those of the
Standard Model and observational cosmology, and to ask whether the
physics of the ensemble of such states is sufficiently similar
that one can make further predictions\cite{douglas}, for as yet
unmeasured quantities\footnote{Another problem here is the
inevitable imprecision of quantities associated with a particular
state in the discretuum, due to its finite width.  For example,
all energy levels should be broadened.  Is this consistent with an
enormously precise tuning of the value of the c.c. associated with
this state?} . At the crudest level, the problem is one of
counting. In order to explain the cosmological constant alone, one
needs of order $10^{63}-10^{120}$ states (depending on one's
assumptions about supersymmetry and supersymmetry breaking).
Additional requirements, such as a suitable hierarchy, the correct
values of the quark and lepton masses and mixings, the correct
amount of inflation, and so on, are likely to increase this by an
enormous factor, quite possibly just as large or larger.  Refererence
\cite{kklt} provided
some estimates of the number of flux vacua, and \cite{douglas}
examined some prototype counting problems, which suggest that
there could conceivably be enough flux vacua.

The problem of actually determining the correct ground state is
hopeless, from this viewpoint.  It is not clear that there is any
sort of small expansion parameter in the flux vacua.  If there is,
it is unlikely to be smaller than a part in $100$. This means
that, if one works say to tenth order in this parameter (far more
precise than any precision QED calculation) one would calculate
the cosmological constant, at best, to an accuracy more than $40$
orders of magnitude less fine than the observed value of the dark
energy. In other words, at this stage of the computation, there
must be {\it at least} of order $10^{40}$ states with properties
otherwise identical to those of the Standard Model; determining
which of these described nature would require an impossibly
difficult calculation.

Instead one would want to
ask whether the bulk of these $10^{40}$ states had some other
characteristic properties, which could lead to predictions for
experiments that were not used as input for determining the
allowed ensemble of states.  We will speculate below that low
energy supersymmetry might be such a prediction.

The alternative mode of explanation is anthropic.  This clearly
raises many of the same issues that we have just discussed.  The
hope, however, is that our presence as observers provides an
explanation of some of the experimental facts which we previously
took as input.  The possibility that such considerations might
explain the value of the cosmological constant has perhaps
received the most attention through the years, but one might hope
to understand many other issues:  the gauge groups of the Standard
Model, the values of some of the quark and lepton masses and the
gauge couplings, and so on.  Within the flux discretuum, as we
will explain in more detail below, all of the parameters of the
Standard Model vary among the different
states. As a result, in this viewpoint, {\it the parameters of the
Standard Model are either anthropic or random}.   As we will
discuss, it is unlikely that even principle A
can explain the values of the Standard Model parameters, or all of
the observed cosmological parameters.  On the other hand the
anthropically unexplained parameters do not appear random, {\it
i.e.} typical of the distributions which seem likely to emerge
from the flux discretuum.  This leaves the possibilities that the
discretuum fails to reproduce the world we see, or that some
additional, rational explanation is required. As we will discuss,
one possibility is that there is a symmetry explanation. For
example, we could imagine that a light $u$ and $d$ quark are
required by anthropic considerations, but that many more states
have light quarks as a result of symmetries than simply by
accident (though crude estimates of the fraction of states in the
discretuum which possess discrete symmetries suggest that this is
not likely to be the case).  

\subsection{An Aside on Continuous Moduli Spaces}

As an illustration of the issues raised in the previous section,
let us first consider a set of states which we have every reason
to believe make sense in quantum gravity.  These are moduli spaces
with more than four supersymmetries, in various numbers of
dimensions greater than or equal to four.  In these cases, the
supersymmetry forbids any potential on the moduli space.  As we
mentioned earlier, string duality has shown convincingly that
there are no non-perturbative difficulties with the construction
of these models.

So what are we to make of these?  In our first paradigm, we would
simply assert that nature is four dimensional, with less than four
supersymmetries.  This is a disappointing retreat, and it is in this
context that the anthropic principle may be necessary and even
appealing. Certainly
carbon-based life will not arise in these states, so principle A
adequate to rule out these
states. It seems quite likely that even principle B
might also successfully explain why life can only exist in a state
with no supersymmetries (possibly four approximate
supersymmetries). Truly establishing this requires careful
thought, but there are likely to be a number of difficulties with
creating any complex structures in such states.
\begin{enumerate}
\item  Conventional stars will not exist, even assuming an
unbroken $U(1)$. \item  Planets, if they exist, will be very
unlike those we see in the universe. \item  It may be difficult,
given the lack of moduli potentials, to understand inflation in
these states.
\end{enumerate}
One can, of course, put forward possible scenarios which would evade each
of these problems.  But it seems to us likely that an application of
principle B might be adequate to understand this most basic fact.

\subsection{Extremely Light Scalars?}

While we are on the subject of the anthropic principle and light
or massless scalars, let us recall a proposal for implementing the
anthropic principle, which involves an extremely light scalar
field\cite{banksls,lindescalar,vilenkin,weinbergls}. The basic idea is very simple.
Suppose, for definiteness, that at very low energies, the
effective lagrangian for this particle is given by: \beq {\cal L}
= {1 \over 2} ((\partial \phi)^2 - m^2 \phi^2) + \Lambda.
\label{verylight} \eeq The time evolution of this field, in an FRW
universe, is given by: \beq \ddot \phi + 3H \dot \phi + m^2 \phi =
0. \eeq If $m \ll H$, then the field $\phi$ is essentially frozen.

Now one imagines that during an extended inflationary period, there are many regions
of the universe with differing values of $\phi$.  If $\phi$ can vary over a huge region
of field space, there will be some regions where the effective cosmological
constant is small enough to satisfy anthropic bounds, and comparable to the
dark energy today.  Note that huge here means that $\phi$ is at least
as large as (assuming that $\Lambda$ in eqn.
\ref{verylight} is of order TeV$^4$)
\beq
\phi = M_p ({M_p^2 \over {\rm TeV}^2}).
\eeq

It is interesting to ask whether fields with such properties exist
in string theory.  Experience with string loop calculations indicates that
conventional ideas about naturalness of scalar
masses are always applicable.  In other words,
it is highly unlikely that one has massless scalars with masses much
less than the scale of supersymmetry breaking (apart from a possible loop factor or two),
except for axion-like objects.

So $\phi$ is most likely an ``axion", i.e. a field with some
periodicity. This axion must have a decay constant enormous
compared to the Planck scale. While no one has proven a theorem
that the Planck scale is an upper bound on axion decay constants,
extensive searches\cite{bdfg} have yielded no examples of
parameterically large decay constants.  So it seems unlikely that
this solution of the cosmological constant problem is
implemented in string theory (the only theory where this question
can be sensibly addressed).

\section{The Anthropic Principle in Flux Compactifications}

In this section, we discuss the anthropic principle as possibly
realized in flux vacua, and its
limitations, in greater detail.

\subsection{The Anthropic Principle as a Datum}

All currently accepted physical theories require some
phenomenological input. Our recent enthusiasm for string theory as
the theory of everything has given rise to the hope that the only
necessary inputs are the basic dimensionful parameters which
define the conversion between socially defined scales of
measurement and the fundamental units of mass, length, time and
action.  This is not necessarily the case, and the existence of
mathematically consistent, disconnected, models of quantum
gravity suggests that it {\it is} not the case.  In this
eventuality, we will have to supply other data as input, in order
to decide on our theory of the world.

A charitable view of the anthropic principle is that it is simply
a novel way of constraining theoretical speculations by fixing a
single complicated piece of data about the world.  The hope of the
anthropically minded is that this single piece of complicated data
will constrain many if not all of the simpler parameters which
appear in the mathematical formulation the theory of the world.

In a previous paper, Motl and two of the authors\cite{bdm}
criticized the use of principle A (the
insistence on carbon based life) as overly parochial.   That
criticism was based on a rather different view of the meaning of
the anthropic principle. It seems tautological that physical
theories which cannot give rise to complex life forms that could
observe their consequences, will never be observed. This
observation seems to make the anthropic principle an axiom,
rather than a radical hypothesis.  However, currently, we cannot
derive the existence of life from fundamental physics.
Furthermore, fundamental physics as we understand it today seems
to allow for the possibility of many theories which resemble the
standard model's gross features (abelian and non-abelian gauge
groups, chiral fermions).  It seems unlikely that we could ever
discover (by pure thought! ) which of these potential low energy
lagrangians could give rise to complex life forms.  If one adopts
only this tautological, but very weak, version of the anthropic
principle (the biothropic principle), it is clear that current
technology can only hope to determine a small number of
parameters by anthropic arguments. Probably the only prediction
of the biothropic principle is a relation between the cosmological
constant, the amplitude of primordial density fluctuations, and
the dark matter density at the beginning of the matter dominated
era, which guarantees that galaxy formation is possible.   Even
``stellar astrophysics" and the details of ``nuclear physics" are
beyond our capabilities for general gauge groups and
representations.

However, from the more modest point of view articulated above,
principle A seems more reasonable.  Indeed, the
attraction of this principle for its advocates is that it appears
to fix many physical parameters close to their observed values. It
can also rule out many of the highly symmetric models of quantum
gravity, with which string theory provides us. The existence of
light photinos with relatively strong coupling, and degenerate
bosonic partners for all fermions, suggests that in an exactly
SUSic theory, the standard model would not give rise to
conventional nuclei and atoms, but rather, Bose condensed clumps
of the superpartners of electrons and nucleons\footnote{Above, we
have suggested that versions of this argument might be valid for a
large class of models of low energy physics.} . In this more
modest view, once we have accepted the existence of the
meta-stable dS states implied by string theory flux
compactifications, it would seem that the
principle A might be an acceptable way to explain why we experience
a particular one of these states.

The key question about principle A thus becomes
whether it is a more economical way of constraining our models of
the world than other pieces of data that one might use to the same
end.  In what follows, we will be examining this question in the
context of flux compactifications of string theory.  The
parameter which cries out for an anthropic explanation in this
context is the cosmological constant, since the meta-stable dS
vacuum energies in these scenarios are not typically small enough
to be compatible with the data.  It was argued by Bousso and
Polchinski (BP)\cite{bp} that with a large enough range of values of a
large enough collection of fluxes, it was natural to have some
meta-stable minima with vacuum energy on the order of that which
is indicated by cosmological observations.   Subsequent work has
tried to demonstrate, with some success, that the requisite
number of vacua and fluxes exists in string compactifications.
One might then argue that anthropic considerations will pick out
a value of the cosmological constant consistent with observation.

We will discuss below the question of whether the anthropic
prediction of the cosmological constant in these models is
successful.  Here we only want to note that if there are any
meta-stable dS solutions consistent with the constraint, there are
likely to be many.  Within the BP ground rules, one would
re-introduce fine tuning if one tried to insist that there were
${\cal O}(1)$ states which satisfied the anthropic
constraint\footnote{This is true even if we consider the
anthropic constraint as a constraint on the cosmological constant
with all other parameters fixed at their observational values. In
the true multiparameter fit, it would require even more tuning.}.
Thus, there will be many minima, with a variety of low energy
effective theories, which satisfy the anthropic constraint on the
cosmological constant.

One of the disappointing things about the
principle A is that many aspects of the choice of low energy gauge
group will be dictated by this principle, rather than determined
dynamically.  If we simply view the principle as a complicated
datum this disappointment might turn into a virtue - our single
piece of input might determine not only the cosmological
constant, but also the structure of the standard model. However,
as we will see in the next few sections, this is unlikely to be
true.   Principle A is not strong enough to
determine the gauge group at the weak scale.

\subsection{The Renormalization Group and the Anthropic Principle}

To understand the utility of the anthropic principle as a piece of
data, we must carefully separate it from other data that have been
used to fix our current understanding of physics.  Anthropic
arguments involve physics at scales of a few hundred MeV or less.
Apart from the parameters in the effective lagrangian at this
scale, they depend on boundary conditions which are determined by
high energy physics.   These are the primordial density
fluctuation spectrum, the baryon asymmetry, the density of dark
matter at the beginning of the matter dominated era, and the fact
of proton stability . Primordial element abundances are not
terribly important, nor are any but the grossest properties of
dark matter necessary to the formation of galaxies.   The elements
that concern us as humans were formed in stars.   For anthropic
purposes, star formation does not have to be such as to reproduce
the observed stellar populations, but merely such as to produce a
reasonable number of stars like our Sun.  The question of what a
reasonable number is constitutes a problem in exobiology which has
not yet been solved.

For anthropic purposes we need only to have a nuclear physics
reasonably like the existing one.  Indeed, most detailed
properties of elements heavier than carbon (or at most iron) are
not important.  They must only be such as to supply a level of
radioactivity sufficient to fuel geological processes, and perhaps
to catalyze an appropriate mutation rate for DNA.

Within the context of string theory, it is plausible to assume
that the origin of nuclei comes from a strong gauge interaction,
but there are lots of possibilities for obtaining particles like
the proton and neutron.  The simplest of these is the $SU(2N+1)$
generalization of QCD with two light flavors.  Note that the
heavier families of quarks and leptons do not participate in
anthropic considerations\footnote{S. Thomas has noted a possible
exception.  A muon with just the right mass gives us cosmic
radiation which penetrates the atmosphere.  One might speculate
that this is important for evolution.   On the other hand - given
our lack of understanding of the value of the muon mass, one can
easily imagine other models which give rise to a particle with the
necessary properties.}.   It is only the crucial dependence of
stellar properties on the marginal stability of deuterium (and
perhaps a few other poorly understood nuclear facts) that might
lead one to believe that the value of $N$ was crucially important.

Nuclear physics at the level of anthropic precision, depends even
less on the weak interactions.   They enter only as four fermi
operators and it is unlikely that even the $V - A$ structure is
crucial.  It is hard to decide this, because our ability to
calculate nuclear properties is limited, but the fact that it took
twenty years, and high precision experiments to establish the $V -
A$ theory suggests strongly that the existence and gross
properties of carbon, nitrogen, oxygen and iron nuclei would not
be changed if the weak interactions were slightly different.

This is even more evident for the neutral current weak
interactions.  Their importance to anthropic arguments is limited
to their role in supernova explosions.  However, to the extent we
understand this, a wide range of additional neutrino interactions
would serve the same purpose.

More generally, there is a wide variety of modifications of the
standard model of electroweak interactions, with new particles at
the $50 - 100$ GeV scale, {\it all of which are both ruled out by
experiment, but compatible with the anthropic principle}.  These
include models which can be derived from string theory {\it e.g.}
intersecting $D6$ brane ``standard-like'' models. Here we make the
assumption that the exotic particles in these theories have masses
which make them observable in current experiments but not in
nuclear physics.

Other crucial anthropic features of the standard model are the
proton neutron mass difference and the electron mass \footnote{We
take for granted the fact that we need a photon with the
right value of the fine structure constant.} .  These involve the
least understood part of the standard model, the light quark and
lepton mass matrices.   We will argue below that it is difficult to
understand many features of these mass matrices in the context of flux
compactifications.  Here we only note that it is easy to satisfy the
anthropic constraints on these masses in many extensions of the standard model which
are ruled out by experiment.  Thus anthropic arguments give us no
way to distinguish between such models.

To sum up, the renormalization group implies
that even principle A is not a powerful piece of data for
discriminating between the behavior of different models at energy
scales of order $100$ GeV.  Anthropically interesting physics
depends on the physics at these scales only through a few
effective parameters.  Flux compactifications appear able to
generate a wide variety of gauge groups and matter
representations, as well as a near
continuum of values of couplings. It seems unlikely that even
the requirement of carbon based life
will be able to pick a unique solution from this
large collection.   For example, even if the gauge group
and matter content is correctly selected, it is unlikely
that anthropic considerations determine these parameters
with sufficient accuracy that they are sensitive
to the values of the $c,b$ and $t$ quark
masses.  If this is correct, and flux compactifications
exist, then it appears that string theory, even when coupled with
principle A, does not give definite predictions
for physics at the electroweak scale.  Indeed, it is likely to
give rise to many anthropically allowed models which are already
inconsistent with experiment.  Of course, given our lack of
knowledge of the landscape of flux compactifications, one can
still argue that one {\it might} find only one minimum with the
right value of the c.c. and approximately the right nuclear
physics.   Our observations should thus be viewed as a challenge
to this set of ideas, rather than a no-go theorem.

There appears to be one possible loophole in this argument. Some
parts of nuclear (and consequently stellar) physics appear very
sensitive to small changes.  We are thinking particularly of the
margin stability of deuterium and the proton-neutron mass
difference as low energy properties that dramatically affect
nuclear and stellar physics. It is possible that a unique solution
could be picked out by these sensitive properties. In order to
verify this conjecture we would have to understand the whole
collection of flux compactification models which give a more or
less standard nuclear physics, and calculate the deuteron binding
energy and proton neutron mass difference in each of them.  The
task of the anthropic flux compactifier is not an easy one.

One always emphasizes to one's students that the renormalization
group is not a group but a semi-group.  The burden of this
subsection has been that this well known fact implies that
principle A is not a terribly powerful piece of
data for discriminating physics at the $100$ GeV scale.  We will
try to order our subsequent remarks about the anthropic principle
in approximate ``RG order" .   That is, we will assume that some
combination of anthropic arguments and knowledge of the flux
landscape has successfully reproduced some aspects of physics at
a given energy scale, and then explore the utility of these
arguments for probing physics at higher energies.   Our next step
is to assume that the standard model gauge group has been fixed
by these considerations and ask what further aspects of the
standard model lagrangian will be simply explained by them.

\subsection{Bottom Up Approach to the Phenomenology of the Discretuum}

The discussion of the renormalization group in the previous
section suggests a ``bottom up" approach to the phenomenology of
the discretuum.  Consider, first, the parameters of the effective
lagrangian well below the QCD scale.   At the smallest energy
scales, we have the cosmological constant.  At higher energies, we
have the electron mass and the electromagnetic coupling (and the
proton mass).  We can well imagine that strong selection effects
might choose among these. At still higher energies we encounter
the parameters associated with nuclear and low energy QCD
interactions.  These include the number of colors, the $u$ and $d$
quark masses and the scale of QCD.  It is easy to imagine that
strong anthropic effects would account for the $u$ and $d$ quark
masses.  For example, if the $u$ quark were heavier than the $d$
quark, the mass of the neutron would be lower than that of the
proton.  Similarly, if the $u$ and/or $d$ quarks had their
``natural" values, of the same order as the weak boson masses,
then nuclear physics as we know it would not exist.   This
constitutes an anthropic argument for the small values of these
masses, which are commonly believed to result from softly broken
discrete symmetries.   In the flux vacuum context, this is
welcome, because, as we will see, discrete symmetries are broken at a high scale in
most of these states.  Similarly, the value of the QCD scale
relative to the electron mass might be fixed by anthropic
considerations in molecular physics.

Stellar processes are sensitive to the precise values of the above
mentioned quantities. These are probably necessary to explain the
gauge group of the strong interactions itself. As we have already
remarked, if there are two light quarks, we might well imagine
that nuclear physics is not qualitatively different for different
values of $N$, and that for suitable quark masses it is even
quantitatively almost the same. Only
a very strong version of principle A, presumably related to the properties of deuterium, are
likely to explain why the gauge group is $SU(3)$ and not something
else.

However, there is one other parameter we encounter at this stage,
which even principle A is not likely to
explain:  the QCD angle, $\theta$.  It is hard to see an anthropic
argument that $\theta$ is small, much less bounded by $10^{-9}.$
Moreover, in the flux vacua, there is typically no light axion.
Recall that the complex structure moduli were fixed, in a
supersymmetry conserving fashion, at scales well above the
supersymmetry breaking scale. The radius, $\rho$, was fixed by the
superpotential: \beq W = W_0 + a e^{-b \rho}. \eeq So the real
part of $\rho$, which we might hope to identify as an axion,
in such models, gets a mass of order the scale of supersymmetry
breaking, and is irrelevant to the strong CP problem. In general,
obtaining light axions once supersymmetry is broken poses various
difficulties\cite{bdg}.  It is clear that a ``generic" flux vacuum
does not readily yield an axion.  Whether some special subset
(e.g. with discrete symmetries) might yield one is an interesting
question. Conceivably, the
axion might be an accidental consequence of, say, discrete
symmetries which are also responsible for the smallness of quark
and lepton masses. However, we have already noted that there is a
conflict between flux vacua and discrete symmetries.

Alternative solutions to the strong CP problem include a massless
$u$ quark. Many theorists would argue that this possibility is now
ruled out both by effective chiral lagrangian arguments and
explicit lattice computations. But even if we are willing to
contemplate it, it seems unlikely that this will arise in flux
models, for the same reason that it is hard to understand light
quarks and leptons:  the absence of discrete symmetries in generic
flux vacua.  Alternatively, as in the models of Nelson and
Barr\cite{barrnelson}, one could contemplate that the possibility
of ``spontaneous" CP violation, with $\theta$ small as a
consequence of accidental features of the model. But given that
typical fluxes violate CP, it is hard to imagine that this could
arise, again, in any but a modest subset of these states.

There is only one way that we can imagine resolving this
conundrum.  We have already argued that anthropic reasoning forces
the two lightest quarks to have masses close to their observed
values.  Suppose we were to find that in the vast discretuum, this
was achieved most often in states with appropriate discrete
symmetries (rather than simply randomly).  Note that this is very different from anthropic
arguments for the cosmological constant, or the weak scale.  In
those cases we abandon the traditional search for symmetry
arguments to explain small numbers, and simply throw away all
states which do not satisfy the anthropic bound.  Now we are
hoping that an anthropic constraint can only be satisfied as a
consequence of a symmetry.    We will refer to such a situation
as one in which we have an Anthropically Required Discrete
Symmetry (ARDS). One might then hope that the symmetry could also
explain the strong CP problem. We emphasize that in order to
verify a claim like this, one would have to search through the
$10^{100}$ states of the discretuum and prove that the
overwhelming majority of states, which have light quark masses
appropriate for anthropic needs, also have discrete symmetries
which somehow solve the strong CP problem.
We will shortly examine a toy model which suggests that the fraction
of states which possess discrete symmetries is likely to be very
small (so small quark masses, obtained randomly, might be much
more likely).
Still, while this possibility
seems unlikely to us at the moment, its verification would be a
triumph for the idea of a string landscape.
This problem also exemplifies the kind of labor that would be involved
in extracting predictions from the landscape.

The strong CP problem is our first serious obstacle to the
anthropic selection mechanism. As we increase the energy scale,
however, we encounter others.  In particular, the masses of the
$s$, $c$ and $b$ quarks, and their mixing angles, are all puzzling
from this viewpoint.  A similar statement applies to the $\mu$ and
$\tau$
leptons (we will not attempt to deal with neutrino masses here).
There is no obvious anthropic argument that these should be
small.  In the absence of some sort of approximate symmetry, we
would expect them all to be of order the $W$ mass. We recall that
typical explanations of these numbers rely on symmetries, but that
typical flux vacua do not seem to have symmetries.  We will
comment on this issue further when we discuss proton decay in
supersymmetric theories.   There is also a wide range of similar
problems associated with other rare processes involving violation
of flavor symmetries of quarks and leptons.  These are most
severe if low energy SUSY, or some other criterion, forces us to
consider new physics at scales that are not much higher than the
weak scale.

There are further severe problems associated with proton decay. As
we will explain, the flux discretuum is likely to predict low
energy supersymmetry, so proton decay is potentially a very
serious problem. Arguments based on Principle A, by
themselves, provide at best a limit on the proton lifetime of
order $10^{16}$ years or so\cite{goldhaber}. So one cannot explain
the large value of the proton lifetime from such considerations.
Alternatively, one can, again, ask whether the proton lifetime can
be accounted for by symmetries. But as for the question of light
quark masses, the problem is that most states in the flux
discretuum can be expected to possess no symmetries. Flux vacua that can account for
observed phenomena have many small parameters (inverse fluxes),
and typically have a rich hierarchy of mass scales, as we have
indicated.   There may be many processes which mediate proton
decay at levels compatible with anthropic, but incompatible with
experimental, bounds.   In vacua with low energy SUSY this will
certainly be a problem.

While in the absence of explicit examples, it is difficult to
make a firm statement about the fraction of states which possess discrete
symmetries, it is difficult to see how this fraction could
be substantial.  If one starts with some $F$-theory compactification
with a large discrete symmetry, and a large number of possible
fluxes (as required for a discretuum), one expects that the fluxes
themselves will break the symmetries.  The issues can be illustrated
with a simpler and better understood example, that of the compactification
of Type IIB theory on a $T_6/Z_2$ orientifold\cite{twoborientifold}.
Here, at particular points in the moduli space, one has a large discrete
symmetry, $Z_2^5 \times S_6$.  One can add fluxes, specified by twenty
different integers, in such a way as to fix all of the complex structure
moduli and the dilaton.  Flux configurations break some or all of the discrete
symmetries.  Taking a value for just one of the integers reduces the symmetry
to $S_3 \times S_3 \times Z_2^4$.  Additional fluxes break further symmetry.
In order to preserve a single $Z_2$, half of the $20$ integers must vanish.  By analogy,
if there were, say, $10^{200}$ states, one might expect that only one in $10^{100}$
possess any discrete symmetry at all.

Of course, we might someday exhibit a large set of flux vacua for
which there are discrete symmetries. Alternatively, if the low
energy theory is not supersymmetric, the situation might be
better.  In such a case, we would have to invoke principle A
selection arguments to account for the weak gauge hierarchy.
Proton stability might then be understood if there were truly a
desert up to the GUT scale. However, since no anthropic argument
requires such a desert, and it is far from clear that typical flux
vacua will produce one, it is not clear that we can understand the
experimental bound on the proton lifetime in this way.

Indeed, the best way to solve most of the problems that we have
enumerated is to imagine that anthropic arguments force us to a
situation in which there is a desert up to the Planck or GUT
scale, inhabited only by standard model particles.  If one then
assumed that the anthropic constraints on the light quark masses
required us to have a set of discrete symmetries which solved the
strong CP problem and explained the quark and lepton mass
matrices, we would have a successful and predictive model of
physics based on the discretuum, albeit a model that holds little
excitement for experimentalists.  At the moment, such a scenario
seems unlikely to us, and the amount of theoretical labor involved
in exploring it is daunting.

To summarize, many of the parameters of the standard model do not
have an anthropic explanation, nor do they appear to be random
numbers.  Generic flux vacua do not support discrete symmetries
which would provide a natural explanation for the values of these
parameters.   Barring remarkable accidents which will only be
explained when we understand the detailed spectrum of flux vacua,
it would seem that the generic prediction of models based on
these meta-stable minima and principle A, is
in conflict with experiment.

\subsection{Might the Flux Discretuum Predict Low Energy
Supersymmetry ?}

In the previous section, we have enumerated a number of
phenomenological difficulties connected with the flux vacua.  In
this section, we will ``suspend disbelief", supposing that some
rational solution of these difficulties (symmetries which survive
in a vast number of states, for example) will be found, and turn
to the question:  can the flux discretuum, coupled with anthropic
arguments, lead to a prediction of low energy SUSY.

While the analysis of \cite{kklt} is based on an action which is
supersymmetric with small $\Lambda$, we have seen that in the vast majority of states,
either one has unbroken supersymmetry, with a deep AdS minimum,
or one has large supersymmetry breaking.  Still, assuming
a vast array of states, some number of states have small $W_0$ and
small scale of supersymmetry breaking.  In a fraction of {\it these},
as a result of
random cancellations,
the cosmological constant will be smaller than the scale
expected from supersymmetry breaking. Indeed, we have seen
that the probability of finding states with small $W_0$ goes as $W_0/M_p^3$.

The key to a prediction of low energy supersymmetry lies in the
question of the cosmological constant.  In non-supersymmetric
states, we might expect that the probability of a low cosmological
constant, of order $\delta^4$ (in Planck units), would be of order
$\delta^4$.  In supersymmetric states, the probability would be
expected to be much higher.  Roughly the probability of a negative
cosmological constant of order $\delta^4 \sim \vert W_0 \vert^2$
would go as $\vert W_0\vert$! So low cosmological constant is far,
far more probable in supersymmetric states. Now we need to fold
this with the probability of breaking supersymmetry, with small
cosmological constant.  Suppose we take as our model for this
phenomenon the possibility that low energy dynamics breaks
supersymmetry.  We might expect that this occurs in a significant
fraction of states (for this purpose, note that a part in
$10^{10}$ is a significant fraction).  This supersymmetry
breaking will generate a positive contribution to the cosmological
constant\footnote{We emphasize that we are using conventional
effective field theory thinking here.  The actual situation in
quantum gravity might be different.  We are currently studying
this puzzle.}. We might expect, if the breaking is dynamical, that
the distribution of this contribution is flat on a logarithmic
scale. So, say, one in 50 of these states will have cosmological
constant which is positive, and within an order of magnitude of
$W_0$. So among the states with small cosmological constant, the
vast majority will have some approximate supersymmetry. Our crude
estimates suggest that the number of supersymmetric states with
cosmological constant $\Lambda=10^{-120}$ is about 70 orders of
magnitude larger than that of non-supersymmetric states with this small a value of $\Lambda$.

This is not actually what we want.  It predicts that supersymmetry
breaking is of order the scale set by the cosmological constant.
However,  principle A also requires us to look
only at vacua with the right value of the weak scale.  The usual
SUSic solution of the gauge hierarchy problem implies that
the number of approximately supersymmetric states, which have SUSY breaking scale of order
the weak scale ($\sim 1 $ TeV), is much larger than the number of
non-supersymmetric states, which have
the weak scale of this order.

A rough estimate of the ratio of non-supersymmetric to supersymmetric states with
suitable cosmological constant and weak scale might be $10^{-34} \times
10^{-120}= 10^{-154}$.  If we suppose, say, low scale supersymmetry breaking, with
$F$ terms of order $10$ TeV, than a plausible guess for the
fraction of suitable supersymmetric states might be $10^{-10} \times 10^{-28}
\times 10^{-2} \times 10^{-63} = 10^{-103}$.  The first factor
is our crude guess as to the fraction of states with suitable supersymmetry
breaking; the factor $10^{-28}$ accounts for the appropriate
size of $W_o$; the factor of 100 accounts for the fraction which have
suitable supersymmetry breaking scale; and the factor of $10^{-63}$ is the
fraction of states for which accidental cancellations might be expected to give
a suitably small cosmological constant.  Note that, in this picture,
the probability of states with suitable properties {\it decreases} rapidly
with {\it increasing} susy breaking scale (roughly as the scale squared).
Thus, the vast majority of states,
which satisfy principle A for both the weak
scale and the cosmological constant, have SUSY breaking of order
the weak scale. Note again that in this more constrained ensemble, the
preference, other things being equal, for small $W_0$ and the requirement
of accidental cancellation of the remaining cosmological
constant, implies
that SUSY should be broken at low energies, as it is in gauge
mediation.

These arguments suggest that the optimistic desert scenario
described in the penultimate paragraph of the last section is not
realized in the flux landscape.  That is, we have identified a
possible prediction which follows from the combination of the gross
structure of the discretuum and the principle A.
This prediction implies that we are saddled with all of the
problems discussed in that section, such as the problem of proton
decay and the smallness of the $\theta$ angle. One can only hope
that the further requirement of small masses for the $u$ and $d$
quarks provide us with ARDS which solve all of these problems.
Otherwise the anthropic predictions of the landscape will be in
gross contradiction with experiment. On the positive side, note
that, since all moduli are frozen at a rather high scale in the
discretuum, one of the cosmological problems of gauge mediation
is automatically solved.

It is worth noting, however, that the prediction of low energy supersymmetry
breaking could well be an example of the sort of ``rational explanation" we have
discussed.  It is unlikely that any anthropic condition would
explain why the squark and slepton spectrum would exhibit the degree
of degeneracy required by existing data.  But if low energy supersymmetry
breaking, with gauge interactions as the mediator, were generic, than
at least this set of questions might find a rational explanation.

\subsection{Above the TeV Scale: Cosmological Constraints}

For reasons which are ultimately tied to the renormalization
group, anthropic arguments are sensitive to physics above the TeV
scale primarily by virtue of its effect on cosmological history.
Consequently, it is most useful to describe the constraints in
terms of effective cosmological parameters, rather than
Lagrangian parameters.   We will find a reasonably wide range of
cosmological parameters to be consistent with the constraints.
{\it A priori} this suggests that an even wider variety of
effective Lagrangians below the GUT scale will be consistent with
them.

Assuming nuclear and stellar physics parameters to be fixed at
their real world values\footnote{In principle we should let both
nuclear and cosmological parameters vary simultaneously to find
the real anthropic range.  This will imply an even wider scatter
than we find in our current discussion.}, the important
cosmological parameters are the c.c.,$\Lambda$, the dark matter
density at the beginning of the matter dominated era, $\rho_0$,
the normalization, $Q$, of the primordial density fluctuations,
and the baryon asymmetry. Note that in making this statement we
are committing ourselves to a rather particular model for
structure formation, namely Cold Dark Matter with a
Harrison-Zeldovich spectrum of Gaussian primordial density
fluctuations.  We do this for simplicity only. If there exists a
collection of flux vacua which predict the correct nuclear
physics, and a distribution of galaxies generated by hot dark
matter with cosmic string seeded density fluctuations, which are
compatible with the existence of human life, then they should be
included in the anthropically allowed list of flux vacua, even if
they disagree violently with observation. In other words, our
{\it a priori} restrictions can only enhance the possibility that
anthropically chosen flux vacua could provide a predictive
framework for fundamental physics.


Many authors have considered the possibility that anthropic considerations
might explain the values of these parameters.  Most analyses consider
the effect of varying one parameter, while holding the others fixed.
Even rather weak anthropic selection constraints are then quite impressive.
Weinberg's constraint on the cosmological constant resulting from galaxy formation
has the form:
\beq{\Lambda \leq T Q^3 \rho_0},\eeq where $T$ is a number of order one. With
other parameters fixed at their true values, this is close to the
observed value.  From the perspective of the discretuum,
it requires that
the discretuum have typical spacings in cosmological constant much
smaller than $10^{-47}$ GeV$^4$. This is probably the most
stringent constraint on the size of the discretuum. Even
if the characteristic scale of the cosmological constant is
as small as we might imagine -- $(10 ~{\rm TeV})^4$, (as might be
the case in low energy gauge mediation) one still requires of
order $10^{63}$ states, {\it to address this issue alone}.
{\it If} anthropic considerations are relevant,this, as we have
seen, is perhaps
one of the stronger arguments for low energy gauge
mediation.

One might hope to constrain the various parameters here independently.
Rees and Tegmark\cite{rees} have attempted to provide
independent anthropic arguments for the value of $Q$.  They claim
that it is bounded between $10^{-4}$ and $10^{-6}$, and suggest
that Vilenkin's ``principle of mediocrity" might show that most
anthropically allowed solutions had $Q \sim 10^{-5}$ as implied
by observation.
However, Aguirre has argued that there are many regions of the cosmological
parameter space which are compatible with even rather strong selection
criteria; among these, the cosmological parameters can vary by many
orders of magnitude\cite{aguirre}\footnote{We thank A. Aguirre
and J. Primack for discussions of these and related issues.}.  It is conceivable that stronger
constraints, and more extensive simulations, will narrow the possibilities
But it seems likely that they will not, and that typical states in the
discretuum compatible with observers will predict cosmological parameters
quite different than their observed values.

Within the framework of inflation and cold dark matter we have adopted,
we can outline a number of potential problems with observational
cosmology in the flux discretuum.
In particular, we will now argue that $Q$ and $\rho_0$ are
unlikely to be fixed to the same values in all anthropically
allowed histories of the flux discretuum.



There are two leading particle physics candidates for dark
matter, SUSY neutralinos, and axions.  The discretuum freezes all
moduli at a rather high energy scale, and thus does not provide us
with any axion candidates.   There are two problems with SUSic
neutralinos, the first being that in the discretuum there is no
reason to expect a conserved R parity which makes the LSP stable.
Let us lump this with the rest of the puzzles which we hope to
solve with ARDS.  We still have a problem, since we have claimed
that the landscape predicts low energy SUSY breaking, which means
that the LSP is a very light goldstino/gravitino, with rather
strong (1-100 TeV scale) coupling to the rest of the contents of
the universe.  This is not a particularly good CDM candidate.   It would appear
that the generic prediction of the landscape is that dark matter
is warm or hot, some combination of gravitini and neutrini.   This is probably
not consistent with observations.

We can instead proceed by assuming an as yet unidentified CDM candidate, which
will be found to be predicted by the discretuum.   By analogy with
the situation for neutralinos, we anticipate that the physics
determining its relic density will be affected by a large number
of parameters in the TeV scale effective Lagrangian.  As a result,
its
value can easily vary by one or two orders of magnitude as we jump between different flux
vacua with standard model parameters in their allowed ranges.

To discuss the likely values for $Q$ we must choose a model for
the origin of fluctuations.  For simplicity we have chosen
inflation.   This leads to a number of interesting questions
about the discretuum.   First, is inflation natural in this
context?   Second, is inflation anthropically favored?   These
are difficult questions, so we pose only tentative answers.

Typical models found in the inflation literature require fine
tuning in the technical field theory sense.   A class of models
with minimal fine tuning is represented by lagrangians of the form
\beq {\cal L} = G_{ij} (\phi^k /m_P) \nabla \phi^i \nabla \phi^j -
{M^6 \over m_P^2} V(\phi^k / m_P ) ,\eeq where $M \sim 10^{16} $
GeV and $m_P$ is the reduced Planck mass. Lagrangians of this type
arise for bulk moduli in brane world models with SUSY broken on
the branes, in which $M$ is the fundamental Planck scale and $m_P$
is a consequence of some modestly large extra
dimensions\cite{hw,bd,tb} . They require ${\cal O}(1/ N_e)$ fine
tuning, to explain $N_e$ e-foldings of inflation, and predict the
right normalization for density fluctuations.   We have seen that
a large radius for the extra dimensions is a natural phenomenon
in flux compactifications. It is also important that the
semiclassical approximations for inflation and tunneling
calculations are justified in these models by the small ratio
$M/m_P$.   The IIB solutions of the flux discretuum are indeed
brane world scenarios of the above type, and so their moduli are
appropriate inflaton candidates.

However, the interesting question is whether the number of
e-foldings necessary to produce enough galaxies to guarantee the
existence of human life is comparable to the number we need to
explain observations.  Our sense is that this is typically not
the case.   Thus, if say $30$ e-foldings were enough to satisfy
anthropic considerations, it would seem that there were many more
histories of the flux universe compatible with life than were
compatible with our observation of scale invariant correlations
in the CMB over most of the horizon.  We strongly suspect that
the anthropically required number of e-foldings is quite a bit
smaller than that required to fit the data.  In the context of
flux compactifications, and given the fine tuning necessary in
order to get large $N_e$ one would suspect that most inflationary
histories compatible with principle A have the
minimum number of e-foldings.

Note that in the phrase ``histories of the universe" , we include
not only the different basins of attraction compatible with life,
but also the initial conditions for the inflationary trajectory
which comes to rest in a given basin.   The possibility of
tunneling between different meta-stable dS minima ensures that
there will be many different trajectories which access a given
basin of attraction.  Each will give a different number of
e-foldings of inflation and a different value of $Q$.  The values
of $Q$ may differ by as much as an order of magnitude or so.

Rees and Tegmark's bounds on $Q$ do not
appear to be independent of $\Lambda$ and are in fact
evaluated only for $\Lambda = 0$.  Thus, for example, the upper
bound comes from the statement that dense supermassive galaxies
would form, in which collisions would disrupt planetary orbits on
a time-scale too short for our kind of life to evolve.   However,
if $\Lambda$ were larger, then the value of $Q$ at which this
occurs would also be larger, so although this bound may have a
somewhat different functional form than Weinberg's, it has a
similar correlation between large values of $Q$ and $\Lambda$.
Consequently, a uniform distribution in the anthropically allowed
region of these two bounds, would favor large values of $Q$,
$\rho_0$ and $\Lambda$, constrained only by the underlying theory
(the discretuum in our case).   Given the fact that Weinberg's
bound already overshoots the observed value for $\Lambda$, when
$Q$ and $\rho_0$ are given their observed values, we doubt that
even the principle of mediocrity can rescue the situation.
Rather, the anthropic prediction for the three parameters will
simply disagree with observation without further theoretical
input.   We have argued above that the discretuum scenario is
unlikely to supply that extra input.  Of course, there is always
the possibility that further explorations of the discretuum will somehow
constrain the relative values of these parameters, or that stronger
anthropic arguments will fix them to their observed values.  But
both possibilities seem unlikely.

\section{\bf Conclusions}

Our remarks have ranged over a wide range of
issues.  Let us try to put some order to our
observations.  We have argued
that the concept of effective potential has no exact meaning in
theories of quantum gravity.  We believe that it is therefore not
a coincidence that one cannot propose a systematic method of
improving the calculations which suggest the existence of a
discretuum in string theory.   For the case of the AdS
discretuum, a possible hypothesis is that it corresponds to a
large class of three dimensional CFT's and that the effective
potential computation is some sort of approximation to the
problem of finding RG fixed points.   Each point in this
discretuum would be an isolated, self consistent model of quantum
gravity in AdS space.

For meta-stable dS minima, we must solve two problems.   We must
show that the disastrous Big Crunch instability is harmless.  If there are non-vanishing tunneling amplitudes
between any two dS minima, even one non-vanishing amplitude to tunnel to the Big Crunch might pose a problem.   To resolve it, one would either have to find a quantum resolution of the Big Crunch, or prove Susskind's conjecture that the singularities are no more dangerous than those of black holes in asymptotically flat space-time.

Even if this problem is resolved,
we must find the quantum theory of the Big Bang cosmologies in
the Dine-Seiberg region, which are the putative stable endpoint
of meta-stable dS decay.  And we must show that the quantum
observables of this model of gravity indeed contain information
relevant to a hypothetical observer during a particular dS era.
The question of whether these minima of the effective potential
are relevant to real models of quantum gravity is not yet
answered.

Leaving aside these questions of principle, we investigated the
likelihood that the discretuum could lead to correct postdictions
for existing experimental data, or predictions for experiments
not yet done.   We argued that a great deal of {\it a priori} data will
have to be supplied, and followed the suggestion of Susskind that
the principle A might be the most
economical way to package this additional input.  We argued that
the renormalization group suggests a problem with this approach.

In order to correctly select the standard model gauge group, we
will have to show that the discretuum does not contain many models
which reproduce nuclear physics as we know it, but disagree with
experiment at the few $100$ GeV energy scale.  We argued that $SU(3)$ may
be selected as the group of the strong interactions by detailed interplay
between nuclear physics and stellar structure, but many possible weak
interaction theories are likely compatible even with the
principle A.

Assuming an argument which fixes the low energy gauge group to be
the standard model, we found a large number of residual problems.
These included the existence of precisely two heavier generations
of quarks and leptons, and the many parameters in the standard
model which take on values which are not determined
anthropically, but are far from random.   This problem was
exacerbated by our argument that low energy SUSY probably is a
prediction of the discretuum.   The combined anthropic
constraints on the cosmological constant and the weak scale, and
the general structure of the discretuum, favor SUSY breaking at
about the weak energy scale.   The result is that our low energy
models have a large number of parameters that are severely
constrained by experiment, but not nearly so severely by anthropic
arguments.

We suggested one possible resolution of this problem.  Anthropic
arguments favor very light $u$ and $d$ quarks, as well as a small
electron mass.  One might imagine showing that within the
discretuum, these values could be obtained in an overwhelming
more probable way, only in the presence of a large set of
discrete symmetries.  One then hopes that these symmetries will
be enough to explain the otherwise peculiar values of a host of
parameters in the low energy Lagrangian.  We do not see any
indication that this optimistic hypothesis is true.  Furthermore,
just like the problem of predicting the standard model gauge
group,  verifying the hypothesis involves a herculean task. One
must survey the entire set of metastable dS minima with the right
values of the cosmological constant, the weak scale, and the
first generation fermion masses and show that the overwhelming
majority of them also have the correct order of magnitude
predictions for the all of the ``peculiar" parameters in the
supersymmetric standard model.

Our final topic was an analysis of strong selection constraints
on physics above the scale of the standard model.  These
constraints operate through cosmology, so we presented our
analysis in terms of effective cosmological parameters, rather
than the high energy Lagrangian parameters which might determine
them.  The center of our analysis was the anthropic constraint on
the cosmological constant, which is really a relation between
that parameter, the ratio of dark matter to photons, and the
amplitude of primordial fluctuations.   We argued that all of
these were expected to fluctuate as we move around the discretuum.

Our first result was a negative one: the combination of the
freezing of moduli in the discretuum, and our prediction of low
energy SUSY, left us with no apparent candidates for cold dark
matter.  The discretuum, plus the constraints on physics at the
standard model scale, seems to predict a hot dark matter
universe, which is allowed anthropically, but not observationally.
Assuming that, once we had better knowledge of the discretuum, a
CDM candidate would be found in the majority of anthropically
allowed states, we argued that the Weinberg relation between
$\Lambda$, $\rho_0$, and $Q$, was likely to predict expectation
values for these parameters in the anthropically allowed ensemble
of flux vacua, which were inconsistent with observations.

Finally, we noted that, since models with $N_e$ e-foldings of
inflation require fine tunings which are at least of order
$1/N_e$, anthropic arguments suggest that the discretuum will
predict a large probability for models with the smallest number
of e-foldings compatible with life.   If that number is of order
$10$, our own universe might appear somewhat special.  However,
this is much less of a problem than most of the erroneous
predictions of the anthropically constrained landscape.

Certainly, if there are any points in the landscape at all, there are many more
than are accessible to semiclassical
analysis.  Furthermore, if one wants to make a real assault on the
problem of making connections between the discretuum and
experiment, one must find ways to calculate the cosmological
constant in any given vacuum with precision $10^{-120}$. Thus, if
one wants to put the lie to the dismal picture of discretuum
phenomenology that we have painted, one must first address the
question of the precise mathematical formalism to which the
effective potential is an approximation.   The proposal of
Silverstein\cite{silverstein} is a step in this direction for AdS
points in the discretuum.  However, we believe we have made it
clear that this is a very different problem than finding the
mathematical framework for meta-stable dS minima.  At present we
have no clue about how one might go about computing corrections to
the classical predictions for these states, in a systematic
manner. Thus, it appears to us that the attempt to address the
phenomenology of the discretuum at a level deeper than our own
analysis, must first face up to the problems of principle with
which we began our discussion.

 \noindent {
{\bf Acknowledgements:} We thank A. Aguirre, M. Douglas, S. Kachru,
J.Primack, E. Silverstein, L. Susskind, S. Thomas, S. Trivedi and
J. Wells
for discussions. Conversations with A. Aguirre and Willy Fischler were
particularly important in shaping our thinking on many of these
issues.  We appreciate detailed comments from A. Aguirre, S. Kachru, J. Polchinski
and L. Susskind on an early version of the manuscript.}

\noindent
This work supported in part by the U.S.
Department of Energy.


\end{document}